\documentclass[conference]{IEEEtran}

\IEEEoverridecommandlockouts

\title{
  Bitwuzla at the SMT-COMP 2020
}

\author{
\IEEEauthorblockN{Aina Niemetz~\orcidicon{0000-0003-2600-5283}}
\IEEEauthorblockA{
\textit{
Stanford University}\\
}
\and
\IEEEauthorblockN{Mathias Preiner~\orcidicon{0000-0002-7142-6258}}
\IEEEauthorblockA{
\textit{
Stanford University}\\
}
}

\usepackage{amsbsy}
\usepackage{amsmath}
\usepackage{amssymb}
\usepackage{amsthm}
\usepackage{array}
\usepackage{booktabs}
\usepackage[noadjust,nocompress]{cite}
\usepackage{enumitem}
\usepackage{framed}
\usepackage{graphicx}
\usepackage{multirow}
\usepackage{tikz}
\usepackage{xspace}
\usepackage{subcaption}

\newcommand{\bv}{BV\xspace}
\newcommand{\qfbv}{QF\_BV\xspace}
\newcommand{\qfabv}{QF\_ABV\xspace}
\newcommand{\qfaufbv}{QF\_AUFBV\xspace}
\newcommand{\qfufbv}{QF\_UFBV\xspace}
\newcommand{\qffp}{QF\_FP\xspace}
\newcommand{\qfbvfp}{QF\_BVFP\xspace}
\newcommand{\qfuffp}{QF\_UFFP\xspace}
\newcommand{\qfabvfp}{QF\_ABVFP\xspace}

\usepackage{scalerel}
\usepackage{tikz}
\usetikzlibrary{svg.path}
\definecolor{orcidlogocol}{HTML}{A6CE39}
\tikzset{
  orcidlogo/.pic={
    \fill[orcidlogocol] svg{M256,128c0,70.7-57.3,128-128,128C57.3,256,0,198.7,0,128C0,57.3,57.3,0,128,0C198.7,0,256,57.3,256,128z};
    \fill[white] svg{M86.3,186.2H70.9V79.1h15.4v48.4V186.2z}
                 svg{M108.9,79.1h41.6c39.6,0,57,28.3,57,53.6c0,27.5-21.5,53.6-56.8,53.6h-41.8V79.1z M124.3,172.4h24.5c34.9,0,42.9-26.5,42.9-39.7c0-21.5-13.7-39.7-43.7-39.7h-23.7V172.4z}
                 svg{M88.7,56.8c0,5.5-4.5,10.1-10.1,10.1c-5.6,0-10.1-4.6-10.1-10.1c0-5.6,4.5-10.1,10.1-10.1C84.2,46.7,88.7,51.3,88.7,56.8z};
  }
}
\newcommand\orcidicon[1]{\href{https://orcid.org/#1}{\mbox{\scalerel*{
\begin{tikzpicture}[yscale=-1,transform shape]
\pic{orcidlogo};
\end{tikzpicture}
}{|}}}}

\usepackage{hyperref} %<--- Load after everything else
\usepackage{cleveref}

\begin{document}
\maketitle

\thispagestyle{plain}
\pagestyle{plain}

\begin{abstract}
  In this paper, we present Bitwuzla, our Satisfiability Modulo Theories (SMT)
  solver for the theories of bit-vectors, floating-points, arrays and
  uninterpreted functions and their combinations.
  We discuss selected features and provide details of its configuration
  and participation in the 2020 edition of the annual SMT competition.
  %This paper serves as system description for this competition.
\end{abstract}

\vspace{2ex}
\section{Introduction}
Bitwuzla is a Satisfiability Modulo Theories (SMT) solver
for the theories of bit-vectors, floating-points, arrays and
uninterpreted functions and their combinations.
Its name is derived from an Austrian dialect expression that can be translated
as ``someone who tinkers with bits''.
Bitwuzla is the successor of our SMT solver Boolector~\cite{cav18},
which supports bit-vectors, arrays and uninterpreted functions.

Bitwuzla implements a lemmas on demand procedure for logics with arrays and
uninterpreted functions that generalizes the lemmas on demand for arrays
approach from~\cite{BrummayerB09} to non-recursive first-order lambda
terms~\cite{Preiner-phd,PreinerNB13}.
For quantifier-free bit-vectors,
it supports
the classic bit-blasting approach~\cite{KroeningS-book08},
different approaches to local
search~\cite{NiemetzPB15,NiemetzPB16,NiemetzPB17,Niemetz-phd},
and a sequential combination of both.
For floating-point logics, Bitwuzla includes SymFPU~\cite{BrainSS19}, a C++
library of bit-vector encodings of floating-point operations.
It further supports unsat core extraction for all supported quantifier-free logics.

This paper serves as system description for Bitwuzla as entered into the
SMT competition 2020~\cite{smtcomp20}.
Bitwuzla is licenced under the MIT license, and releases and more information
is available on its website~\cite{website}.

\vspace{2ex}
\section{Features}

\vspace{2ex}
\subsection{Arrays and Uninterpreted Functions}

Bitwuzla generalizes the lemmas on demand for extensional
arrays approach~\cite{BrummayerB09} to non-recursive first-order lambda
terms~\cite{Preiner-phd,PreinerNB13}, which enables compact
representations for operations such as memset and memcpy~\cite{PreinerNB15} and
constant arrays.
It further supports dual propagation-based and justification-based optimization
techniques for lemmas on demand, where the overhead for consistency checking
is reduced by extracting partial candidate models via don't care reasoning
on full candidate models~\cite{NiemetzPB14}.

\vspace{2ex}
\subsection{Quantifier-Free Bit-Vectors}

Bitwuzla implements two orthogonal strategies for solving quantifier-free
bit-vector constraints: the classic bit-blasting approach employed by most
state-of-the-art bit-vector solvers, and local search.
Since local search procedures are only able to determine satisfiability,
Bitwuzla allows to combine local search with bit-blasting in a sequential
portfolio setting, where the local search procedure is run until a certain
limit is reached, before falling back to the bit-blasting engine.

\vspace{2ex}
\noindent
\textit{Local Search for Quantifier-Free Bit-Vectors.}
Bitwuzla supports the stochastic local search (SLS) approach presented
in~\cite{FroehlichBWH15},
an improved variant where SLS is augmented with a propagation-based
strategy~\cite{NiemetzPB15},
and mainly the complete propagation-based local search procedure presented
in~\cite{NiemetzPB17}.
The latter can both be applied on the bit-level and word-level.
The word-level strategy, however is superior to the bit-level implementation,
which operates on the circuit representation of the input formula.
Bitwuzla further implements a novel generalization of the propagation-based
approach in~\cite{NiemetzPB17} to ternary values.
This generalization addresses the main weakness of the propagation-based
local search strategy~\cite{NiemetzPB17,Niemetz-phd}, its obliviousness to bits
that can be simplified to constant values.
The local search engines can now also be combined with the lemmas on demand
engine and quantified bit-vectors.

\vspace{2ex}
\noindent
\textit{Bit-Blasting.}
Bitwuzla implements bit-blasting in two phases.
Initially, it generates an
And-Inverter Graph (AIG) circuit representation of the simplified
input formula and then applies AIG-level
rewriting~\cite{BrummayerB06}.
The rewritten AIG representation is then converted into Conjunctive Normal Form
(CNF) and sent to one of following SAT back ends:
MiniSat~\cite{EenS03},
PicoSAT~\cite{Biere08},
Lingeling~\cite{Biere-SAT-Competition-2018-solvers},
CaDiCaL~\cite{Biere-SAT-Race-2019-solvers},
CryptoMiniSat~\cite{SoosNC09},
or
Kissat~\cite{kissat}.

Bitwuzla uses CaDiCaL version 1.2.1 as default SAT back end.
It further utilizes Lingeling for preprocessing the Boolean skeleton of the
input formula.

\subsection{Quantified Bit-Vectors}
Bitwuzla implements a combination of counterexample guided quantifier
instantiation and syntax-guided synthesis (SyGuS)~\cite{AlurBJMRSSSTU13}
to synthesize Skolem functions~\cite{PreinerNB-tacas17} for existential
variables.
By default, Bitwuzla also employs a dual approach, which applies the same
technique to the negation of the input formula to synthesize quantifier
instantiations.
Both approaches are run in two separate threads in parallel.
Combination with other theories, incremental solving and unsat core extraction
is currently not supported for quantified bit-vectors.

\subsection{Floating-Points}
For the theory of floating-points, Bitwuzla implements an eager translation
of the simplified input formula to the theory of bit-vectors.
This approach is sometimes also referred to as \emph{word-blasting}.
To translate floating-point expressions to the word-level,
Bitwuzla integrates SymFPU~\cite{BrainSS19}, a C++ library of bit-vector
encodings of floating-point operations.
SymFPU uses templated types for Booleans, (un)signed bit-vectors, rounding
modes and floating-point formats, which allows to plug it in as a back end
while utilizing solver-specific representations.
It is also integrated in the SMT solver CVC4~\cite{cvc4}.

\subsection{Unsat Cores}
Bitwuzla implements unsat core extraction via solving under
assumptions~\cite{EenS03}.
When unsat core extraction is enabled, all assertions in the formula are
assumed in the SAT back end.
If given input formula is unsatisfiable, Bitwuzla returns all unsatisfiable
(failed) assumptions as unsat core.
Unsat Core extraction is not yet supported for quantified formulas.

\section{Configurations}

Bitwuzla participates in the single query, incremental, unsat core, and
model validation tracks in the following divisions:

\begin{itemize}
  \vspace{1ex}
  \item \textit{Single Query Track (SQ):}\\
  \bv, \qfbv, \qfabv, \qfaufbv, \qfufbv, \qffp, \qfbvfp, \qfabvfp, \qfuffp
  \vspace{2ex}
  \item \textit{Incremental Track (INC):}\\
  \qfbv, \qfabv, \qfaufbv, \qfufbv, \qffp, \qfbvfp, \qfabvfp, \qfuffp
  \vspace{2ex}
  \item \textit{Unsat Core Track (UC):}\\
  \qfbv, \qfabv, \qfaufbv, \qfufbv, \qffp, \qfbvfp, \qfabvfp, \qfuffp
  \vspace{2ex}
  \item \textit{Model Validation Track (MV):}\\
  \qfbv
\end{itemize}

\vspace{2ex}
\noindent
For divisions \bv and \qfbv in the SQ and MV track,
Bitwuzla uses a sequential combination of bit-blasting and
propagation-based local search with a limit of 10k propagation steps and 2M
model update steps.

\section{License}
Bitwuzla is licensed under the MIT license.
For more details, refer to the actual license text, which is distributed with
the source code.

\pagebreak
\bibliographystyle{abbrv}
\bibliography{biblio}
\end{document}